\documentclass[12pt, a4paper, oneside]{article}

\usepackage{bm} % Bold in math
  \usepackage{natbib}  % or another BibTeX-compatible citation package
  \usepackage{xcolor, soul} % Coloring
\sethlcolor{lightgray}  %Color in highlights in tables
\definecolor{myred}{RGB}{128,22,56} % Define red
\usepackage[colorlinks = true,
    linkcolor = myred,
    anchorcolor = myred,
    citecolor = myred,
    filecolor = myred,
    urlcolor = myred]{hyperref} % Bibliography coloring

\usepackage[margin=1in]{geometry} % make page margins 1inch 
\usepackage{multirow} 
\usepackage{comment}
\usepackage{subcaption}  % For subfigures

\usepackage{amsmath, amssymb} % adds \eqref command
\usepackage{algorithm} % Define Algorithms
\usepackage{algpseudocode} % Define Algorithmics - end, for, state
\algdef{SE}[SUBALG]{Indent}{EndIndent}{}{\algorithmicend\ } % Indent
\algtext*{Indent} % Indent
\algtext*{EndIndent} % Indent
\usepackage{graphicx} % Include figures

\usepackage{tikz} %graph drawing
\usetikzlibrary{shapes,arrows}
\tikzstyle{block} = [rectangle, draw,
    text width=5em, text centered, rounded corners, minimum height=4em]
\tikzstyle{line} = [draw, -latex']    
\tikzstyle{optional}=[dashed,fill=gray!50]
\tikzstyle{VineNode} = [ellipse, fill = white, draw = black, text = black, align = center, minimum height = 1cm, minimum width = 1cm]
\tikzstyle{DummyNode}  = [draw = none, fill = none, text = black] 
\tikzstyle{TreeLabels} = [draw = none, fill = none, text = black] % T_1, T_2, etc.
\newcommand{\xshiftNodes}{0.7*\linewidth}
\newcommand{\yshiftLabels}{-.25cm}  
\newcommand{\labelsize}{\footnotesize} 
\newcommand{\xshiftTree}{0.5cm}    % Distance between trees

\usepackage{xfrac}

\usepackage{sectsty}
\usepackage{color}

\subsubsectionfont{\normalsize}
\subsectionfont{\normalsize}
\sectionfont{\normalsize}

\usepackage[autostyle, english = american]{csquotes}
\MakeOuterQuote{"}
\usepackage{graphics}

\begin{document}

\date{}
\title{Probabilistic patient risk profiling with pair-copula constructions}

\author{{\"O}zge \c{S}ahin\footnote{Corresponding author: O.Sahin@tudelft.nl}\\ \footnotesize Delft Institute of Applied Mathematics, Delft University of Technology, The Netherlands 
}
\maketitle

%TC:ignore
\begin{abstract}
We propose vine copula-based classifiers for probabilistic risk prediction in perioperative settings. We obtain full joint probability models for mixed continuous-ordinal variables by fitting a separate vine copula to each outcome class, capturing nonlinear and tail-asymmetric dependence. In a cohort of 767 elective bowel surgeries (81 serious vs.\ 686 non-serious complications), posterior probabilities from the fitted vine classification models are used to allocate patients into low-, moderate-, and high-risk groups. Compared to weighted logistic regression and random forests with stratified sampling, the vine copula-based classifiers achieve up to 10\% lower class-specific Brier scores and negative log-likelihoods on the out-of-sample. The vine copula-based classifier identifies a large cohort of true low-risk patients potentially eligible for early discharge. Scenario analyses based on the fitted vine copula models provide interpretable risk profiles, including nonlinear relationships between body mass index, surgery duration, and blood loss, which might remain undetected under linear models. These results demonstrate that vine copula-based classifiers offer a reliable and interpretable framework for individualized, probability-based patient risk profiling. As such, they represent a new, promising tool for data-driven decision-making in perioperative care. 
\end{abstract}
%TC:endignore

%\noindent
%{\footnotesize Acknowledgements: We thank Reini Bretveld for facilitating access to the data used in this study. This work is part of the 4TU programme RECENTRE (Risk-based lifEstyle Change: daily-lifE moNiToring and REcommendations). RECENTRE is funded by the 4TU programme High Tech for a Sustainable Future. 4TU is the federation of the four technical universities in the Netherlands (Delft University of Technology, Eindhoven University of Technology, University of Twente, and Wageningen University and Research). 

%Declaration of conflicting interest: All authors have none to declare.
%}%

\newpage

\section{Introduction}\label{sec:intro}
A probabilistic assessment of patients' perioperative risk of having adverse outcomes is critical for clinical decision-making, resource allocation, and individualized care. Given a random vector of clinical variables $\mathbf X$ and the corresponding realization $\mathbf x$, one goal is to estimate $\Pr(Y = 1 \mid \mathbf X = \mathbf x)$, where $Y \in \{0,1\}$ denotes the occurrence of a serious adverse outcome, and translate these probabilities into actionable categories such as low-, moderate-, and high-risk groups. While the low-risk group can be discharged early from the hospital, the high-risk group might need close postoperative monitoring. For the estimation goal, logistic regression remains widely used for its interpretability, but its assumptions of linear log-odds can lead to biased estimates, particularly in the presence of nonlinearity, tail dependence, or mixed continuous-ordinal variables. While tree-based methods can capture variable interactions and nonlinearity, their probabilistic predictions are often miscalibrated in sparse regions of the variable space, which limits their use for reliable probabilistic risk grouping.

Copula models address these limitations by separating marginal behavior from multivariate dependence \citep{Joe2014}. In particular, pair-copula constructions (or vine copulas) offer a highly flexible model that can capture nonlinear, tail-asymmetric dependence observed in the mixed types of variables (continuous-ordinal) \citep{Joe2014, panagiotelis2012pair, Czado2019}. Even though copulas are increasingly used in survival analysis \citep{pate2023developing}, their application to clinical risk prediction remains scarce. Existing approaches embed copulas into logistic regression models \citep{black2021development}, derive classification rules from bivariate copulas \citep{islam2020copula}, or use copulas solely for dependence modeling \citep{kim2025integrating,khosheghbal2025mechanistic}. To our knowledge, no study has utilized class-specific vine copulas (vine copula-based classifiers) proposed by \cite{sahin2024vine} to estimate posterior probabilities for patient risk profiling.

We apply vine copula-based classifiers to a cohort of 767 patients who underwent elective bowel surgery at Medisch Spectrum Twente (Netherlands, 03/2020-12/2023), where the goal is to predict the occurrence of serious postoperative complications, defined by a Clavien-Dindo score of $\ge \text{IIIa}$ \citep{clavien1992proposed}. Our variable set includes three continuous variables of a patient, body mass index, surgery duration in minutes, and age, and two ordinal variables of a patient, each with three levels, blood loss categories and crystalloids administration categories. We train the model on 580 patients (518 non-serious, 62 serious; data from March 2020 to December 2022), and validate it on 187 patients (168 non-serious, 19 serious; data from January 2023 to December 2023). Due to class imbalance, we assess model performance using class-specific Brier scores and negative log-likelihood metrics, which are more informative for probability estimation than the area under the receiver operating characteristic curve (AUC).

Our contributions are: (\emph{i}) we adapt the generative vine copula classifier to mixed continuous–ordinal predictors, benchmark its performance, and provide diagnostics for the simplifying assumption; (\emph{ii}) we propose a probabilistic patient risk profiling framework using class-specific posterior probabilities and decision thresholds, giving clinically actionable low-, moderate-, and high-risk groups; and (\emph{iii}) we use the fitted vine models to generate two-dimensional risk surfaces that visualize nonlinear interactions and provide clinicians with interpretable “what-if” scenarios for preoperative counseling. 

Our results show that vine copula-based classifiers outperform weighted logistic regression and random forests with stratified sampling in both in- and out-of-sample data. They are particularly accurate in identifying low-risk patients, which is an important consideration for early discharge and postoperative care planning. Scenario-based analyses derived from the fitted vine copula models reveal a nonlinear relationship between bleeding volume and surgery duration. 
%To sum up, we highlight the potential of vine-based approaches as reliable, interpretable tools for clinical risk prediction.

The paper is structured as follows. Section~\ref{sec:cop} provides background on copulas and pair-copula constructions, while Section~\ref{sec:vineclass} details vine copula-based classifiers and probabilistic classification measures and presents the proposed risk groups as a probabilistic patient risk profiling framework. In Section~\ref{sec:data}, we present data, the in-sample and out-of-sample results, interpretation of the fitted vine copula models, diagnostic checks, identification of patient risk groups, and scenario analyses. Section~\ref{sec:disc} concludes with implications, limitations, and directions for future research.

\section{Copulas and pair-copula constructions}\label{sec:cop}
In this section, we review key concepts in copula-based modeling by using a clinically motivated example. We are interested in the joint distribution of body mass index (BMI) and surgery duration for patients undergoing bowel surgeries. Let us denote these continuous random variables by $(X_1, X_2)$. We aim to understand their dependence structure, which may extend to a multivariate setting $(X_1,\dots, X_d)$ by including further clinical variables (e.g., age, blood loss).

Each variable has its own \emph{marginal} distribution that captures its univariate characteristics. For example, BMI might be skewed if many patients are obese. Second, we often want to understand how these variables relate beyond univariate characteristics. For instance, we want to analyze if relationships among variables are nonlinear or have tail dependence (e.g., severely obese patients having extremely long surgeries). \emph{Copulas} enable the modeling of such joint dependence separately from marginal distributions. 
%Thus, they are more flexible than other multivariate distributions like the Gaussian. 
%We provide illustrative bivariate examples in \autoref{sec:app-copulas}. 

%\subsubsection*{Sklar's theorem}

Formally, \cite{Sklar1959} showed that for a $d$-variate distribution $F \in \mathcal{F}(F_1,\ldots,F_d)$ with univariate margins $F_j(.)$ for $j=1,\ldots,d$, there exists a distribution function $C:[0,1]^d \rightarrow [0,1]$ (the copula) having $\mathrm{U}(0,1)$ margins, such that
\begin{align}
    F(\bm{x}) 
    \;=\; 
    C\bigl(F_1(x_1),\dots,F_d(x_d)\bigr), 
    \quad \bm{x}\in\mathbb{R}^d.
    \label{eq:Sklar}
\end{align}
Once each margin $F_j(.)$ is specified, $F(.)$ is obtained by plugging in these margins into $C(.)$. If $F(.)$ is continuous with quantile functions $F_1^{-1}(.), \dots, F_d^{-1}(.)$, the inverse of Sklar’s theorem ensures
$
    C(\bm{u})
    \;=\;
    F\!\bigl(
        F_1^{-1}(u_1),\dots,F_d^{-1}(u_d)
    \bigr), 
    \quad
    \bm{u}\in [0,1]^d$,
is unique. Thus, if all margins are continuous, the decomposition in Equation \eqref{eq:Sklar} is unique. When some variables are discrete, the copula is unique over the set $\mathrm{Range}(F_1)\times \cdots \times \mathrm{Range}(F_d)$, but copula-based modeling can be used for data applications. \citep{genest2007primer}. 

Differentiating Equation \eqref{eq:Sklar} with respect to $\bm{x}$ results in the joint density as follows:
\begin{align}
    f(\bm{x})
    \;=\;
    c\bigl(F_1(x_1),\dots,F_d(x_d)\bigr)
    \,\times\,
    \prod_{j=1}^d f_j(x_j),
    \quad \bm{x}\in\mathbb{R}^d,
    \label{eq:Sklar-dens}
\end{align}
where $c(\cdot)$ is the copula density and $f_j(\cdot)$ the marginal density of variable $j$. In a bivariate case, one can also obtain the conditional density as 
$f(x_1 \,\mid\, x_2)
\;=\;
c_{12}\bigl(F_1(x_1),\,F_2(x_2)\bigr)\,\times\,f_1(x_1),
$
which can be interpretable as "Given that a patient is of age $x_2$, what is the distribution of BMI $x_1$?" in a way that allows for tail dependence and nonlinearity, for instance. In higher dimensions, conditional densities can be obtained similarly:
$
f\bigl(x \mid \mathbf{v}\bigr) 
=
c_{XV_j \mid \mathbf{V_{-j}}}\!\Bigl(F\bigl(x \mid \mathbf{v_{-j}}\bigr),\,F\bigl(v_j \mid \mathbf{v_{-j}}\bigr); \mathbf{v_{-j}}\Bigr)\,\times\,f\bigl(x \mid \mathbf{v_{-j}}\bigr),
$
where $c_{XV_j \mid \mathbf{V_{-j}}}(.)$ captures the dependence between the random variables $X$ and $V_j$ conditioned on the remaining variables $\mathbf{V_{-j}}$. Recursively applying these conditional relationships \citep{Joe1996} provides a fully specified joint distribution.

Common parametric copula families include the multivariate Gaussian and the Student-$t$ copula. 
%The Gaussian copula with correlation matrix $\bm{\Sigma}$ is given by $ C(\bm{u}; \bm{\Sigma})
%\;=\;
%\Phi_d\!\Bigl(\Phi^{-1}(u_1),\dots,\Phi^{-1}(u_d);\,\bm{\Sigma}\Bigr),
%$
%where $\Phi_d(.)$ is the $d$-variate normal CDF and $\Phi^{-1}(.)$ is the univariate normal quantile function. A $t$ copula adds a tail-dependence parameter, allowing heavier joint tails than the Gaussian. In the BMI-surgery duration example, a Gaussian copula may suffice if their bivariate dependence is linear (plus noise). However, if extremes (e.g., severely obese patients undergoing very long operations) are more likely, a $t$ copula can capture such symmetric upper-tail and lower-tail dependence.
Asymmetric dependencies (e.g., upper-tail dependence only) cannot be modeled by Gaussian or $t$ copulas. In such cases, Archimedean copulas, defined by a generator function, are often used \citep{genest1986,nelsen1999}. 
%A bivariate Archimedean copula has the form $
%    C_{\varphi}(u_1,u_2)
%    \;=\;
%    \varphi^{-1}\!\bigl(\varphi(u_1)+\varphi(u_2)\bigr),$
%where $\varphi:[0,1]\to[0,\infty)$ is continuous, convex, strictly decreasing, and satisfies $\varphi(0)=\infty$, $\varphi(1)=0$. For instance, $\varphi(x;\delta)=\bigl[-\log x\bigr]^\delta$ with $\delta\in [1,\infty)$ gives the Gumbel copula, which models upper-tail dependence. Other families (Clayton, Joe, Frank, BB1, BB6, BB7, BB8) cover different tail behaviors and can be extended or rotated to allow negative dependence \citep{Czado2019}. 
Bivariate Archimedean copulas typically involve one or two parameters controlling the dependence strength and tail structure. Many families admit a one-to-one mapping between their parameters and Kendall’s $\tau$ \citep{Kendall1938}, a rank-based monotonic association measure. Table 3.2 of \cite{Czado2019} details these relationships.

While Gaussian, Student-$t$, and Archimedean copulas are flexible in two dimensions, they can become restrictive in higher dimensions, especially if the dependence structure varies across different pairs of variables. 
%For example, in three dimensions, if BMI and surgery duration show upper-tail dependence, but age does not share that same tail behavior with BMI, they fail to capture these asymmetric pairwise dependence jointly.
\emph{Pair-copula constructions} (often called \emph{vine copulas}) avoid these limitations by gluing bivariate copulas to have flexible multivariate distribution functions \citep{Joe1996, Bedford2001, Bedford2002, Aas2009}. This approach allows each pair of variables to have its own copula, thereby accommodating diverse dependence patterns among variables.

Their joint density in $d$ dimensions can be written explicitly in terms of $\frac{d\,(d-1)}{2}$ bivariate copula densities. Consider three continuous variables: BMI ($X_1$), surgery duration ($X_2$), and age ($X_3$). By Sklar’s theorem and the chain rule of conditional densities, we can write their joint density $f_{123}(\bm{x})$ as:
\footnotesize
\begin{equation}
\begin{aligned}
f_{123}(\bm{x})
&=\; f_1(x_1)\,\times\,f_{23\mid 1}(x_2, x_3 \mid x_1)
\\[2pt]
&=\; f_1(x_1)\,\times\,
    c_{23;1}\!\Bigl(F_{2\mid1}(x_2\mid x_1),\,F_{3\mid1}(x_3\mid x_1)\,;\,x_1\Bigr)
    \,\times\,
    f_{2\mid1}(x_2\mid x_1)\,\times\,f_{3\mid1}(x_3\mid x_1)
\\[2pt]
&=\;
    c_{12}\!\Bigl(F_1(x_1),\,F_2(x_2)\Bigr)
    \;\times\;
    c_{13}\!\Bigl(F_1(x_1),\,F_3(x_3)\Bigr)
    \;\times\;
    c_{23;1}\!\Bigl(F_{2\mid1}(x_2\mid x_1),\,F_{3\mid1}(x_3\mid x_1)\,;\,x_1\Bigr)
\\
&\quad\quad\quad
    \times\,
    f_1(x_1)\,f_2(x_2)\,f_3(x_3),
\end{aligned}
\label{eq:decomp}
\end{equation}
\normalsize
where $f_j(\cdot)$ are the marginal densities for $X_j$ for $j=1,2,3$, $c_{12}(.)$ and $c_{13}(.)$ are unconditional bivariate copulas modeling the unconditional dependence of $(X_1,X_2)$ and $(X_1,X_3)$, and $c_{23;1}(.)$ is the pair-copula modeling the conditional dependence of  $(X_2,X_3)$ given $X_1$. %As shown in Equation \eqref{eq:decomp}, the joint density can be decomposed as a product of marginal and pair-copula densities.
If instead one decomposes $f_{123}(\bm{x})$ as $f_2(x_2)\times f_{13\mid2}(x_1,x_3\mid x_2)$ in Equation \eqref{eq:decomp}, the set of pair-copulas in the new decomposition differs compared to the ones appearing in Equation \eqref{eq:decomp} (except for $c_{12}(.)$). Thus, multiple ways exist to perform the decomposition, motivating the concept of a \emph{vine structure}.

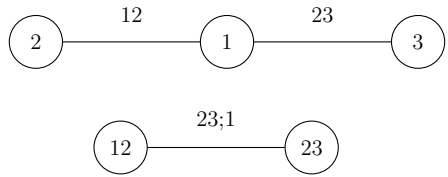
\begin{figure}[ht!]
	\centering
    \caption{An example of a 3-dimensional vine tree structure. Each edge corresponds to a pair copula in Equation~\eqref{eq:decomp}.}
\label{fig:3Dvine}
	\renewcommand{\xshiftNodes}{0.1*\linewidth}
	\renewcommand{\yshiftLabels}{.0cm}  
	\renewcommand{\labelsize}{\normalsize} 
\begin{tikzpicture}[every node/.style = VineNode, node distance =2cm, scale=0.5, transform shape]
\node (v1) {2}			
node  (v2) [right of=v1, xshift=\xshiftNodes] {1}
node  (v3) [right of=v2, xshift=\xshiftNodes] {3}
node  (v4) [below of=v1, xshift=\xshiftNodes] {12}
node  (v5) [right of=v4, xshift=\xshiftNodes] {23};
\draw[color=black] (v1) -- node[draw=none, font=\labelsize, fill=none, above, yshift=\yshiftLabels]{12} (v2);
\draw[color=black] (v2) -- node[draw=none, font=\labelsize, fill=none, above, yshift=\yshiftLabels]{23} (v3);
\draw[color=black] (v4) -- node[draw=none, font=\labelsize, fill=none, above, yshift=\yshiftLabels]{23;1} (v5);
\end{tikzpicture}
\end{figure}

The term \emph{vine} arises from showing the pair-copula constructions via a layered (tree) graph \citep{Bedford2002}, where each edge corresponds to a bivariate copula. Figure~\ref{fig:3Dvine} illustrates a 3-dimensional vine, reflecting the decomposition in Equation \eqref{eq:decomp}. For three variables, there are two tree levels; $d$-dimensional settings have $d-1$ tree levels. The edge $12$ models unconditional dependence of $(X_1,X_2)$ by the bivariate copula $c_{12}(.)$, whereas the edge $23;1$ represents the conditional dependence of  $(X_2,X_3)$ given $X_1$ modeled by the pair-copula $c_{23;1}(.)$.

%\subsubsection*{General $d$-dimensional vine copula constructions}

These ideas generalize to any dimension $d$. A $d$-variate density can be organized across $d-1$ tree levels in a vine tree $\mathcal{V}$. Let $c_{e_a,e_b;D_e}(.)$ denote a pair-copula density on edge $e$ and $\bm{\theta}_{e_a,e_b;D_e}$ its parameter vector. Let $f_p(.)$ be the (parametric or nonparametric) marginal density of variable $X_p$. Then, the joint density $f(\bm{x};\,\bm{\psi})$ can be written as
\begin{equation}
\label{eq:joint-density-d-simplified}
\begin{aligned}
f(\bm{x};\,\bm{\psi})
\;=\;
\Bigl(\!\prod_{m=1}^{d-1}\!\;\prod_{e \in E_{m}}\!
&c_{\,e_a,e_b;\,D_e}\Bigl(
  F_{\,e_a\mid D_e}(x_{\,e_a}\mid\bm{x}_{D_e}),\,
  F_{\,e_b\mid D_e}(x_{\,e_b}\mid\bm{x}_{D_e});
  \;\bm{x}_{D_e},\bm{\theta}_{\,e_a,e_b;\,D_e}
\Bigr)\Bigr)
\\[-2pt]
&\;\times\;\prod_{p=1}^{d} f_{p}\bigl(x_{p}\bigr),
\end{aligned}
\end{equation}
where $\bm{x}=(x_1,\dots,x_d)^\top$, $D_e$ is the conditioning set of size $m-1$ at tree level $m$, and $e_a,e_b$ denote the conditioned variables on edge $e$. The parameter vector $\bm{\psi}$ has all copula parameters $\bm{\theta}_{e_a,e_b;D_e}$ and any marginal parameters.

In practice, a simplifying assumption sets that each conditional pair-copula $c_{\,e_a,e_b;\,D_e}(.)$ in Equation \eqref{eq:joint-density-d-simplified} does not depend on the specific conditioning values $\bm{x}_{D_e}$. This reduces computational complexity and simplifies interpretation, and there are methods to check if it holds in applications \citep{kurz2022testing, chang2020copula}. One may also use truncated vines, assuming independence for edges beyond a certain tree level $m$ \citep{brechmann2012truncated}. For instance, in Figure~\ref{fig:3Dvine}, assuming $c_{23;1(.)}$ is an independence copula yields a 1-truncated vine where one assumes conditional independence of $(X_2,X_3)$ given $X_1$.

Vine copulas can also be used to decompose multivariate probability mass functions (pmfs) when all variables \( X_1, \dots, X_d \) are discrete \citep{panagiotelis2012pair}. Analogous to the continuous case in Equation~\eqref{eq:joint-density-d-simplified}, the joint pmf can be factorized into successive conditional pmfs. The key distinction with  Equation~\eqref{eq:joint-density-d-simplified} is that partial derivatives in the continuous case are replaced by finite-difference approximations in the discrete case.

%For instance, a conditional probability can be written as:
%\footnotesize
%\begin{equation}\label{eq: discrete conditional density}
%    \mathbb{P}(x \mid \mathbf{v}) 
%    \;=\;
%    \frac{\mathbb{P}(x, v_j \mid \mathbf{v_{-j}})}{\mathbb{P}(v_j \mid \mathbf{v_{-j}})}
%    \;=\;
%    \frac{\displaystyle \sum_{i_1=0}^1 \sum_{i_2=0}^1 (-1)^{i_1 + i_2}
%     \,C_{\,X,\,V_j \mid \mathbf{V_{-j}}}\Bigl(F_{\,X\mid \mathbf{V_{-j}}}%(x{-}i_1\mid\mathbf{v_{-j}}),
%     F_{\,V_j\mid \mathbf{V_{-j}}}(v_j{-}i_2\mid \mathbf{v_{-j}});\mathbf{v_{-j}}\Bigr)}{\mathbb{P}(v_j \mid \mathbf{v_{-j}})}.
%\end{equation}
%\normalsize

%In practice, when all variables on an edge of a vine are binary, conditional and joint distributions can be estimated directly from observed frequencies without fitting a bivariate copula. If only some variables are binary, copula families can still be fitted using the subset of non-degenerate data.
The vine copula decomposition naturally extends to data sets with mixed continuous-discrete variables by combining continuous and discrete pair-copula building blocks. We refer to \cite{chang2019prediction} for more details.

In practical applications, (vine) copula estimation often follows a two-stage approach: (i) fit each marginal (parametrically or nonparametrically), and (ii) estimate copula parameters by maximum likelihood \citep{joe2005asymptotic,genest1995semiparametric}. 

In addition, for higher dimensions, there are multiple ways to build the vine decomposition (the choice of tree structure, pair-copula families, and pair-copula parameters). This can be approached via heuristic algorithms \citep{brechmann2012truncated, panagiotelis2012pair}.  %We aim to balance model flexibility with parsimony for clinical applications with a modest number of observations.

\section{Methodology: Vine copula-based classifiers \& probabilistic patient risk profiling}\label{sec:vineclass}
\subsection{Vine copula-based classifiers}
In many clinical applications, one aims to predict a categorical outcome \(Y \in \{1, \ldots, K\}\) (for example, the occurrence of a postoperative complication), with prior probabilities \(\pi_j = \Pr(Y = j)\), based on a vector of mixed discrete-continuous variables \(\mathbf{X}\). Thus, the goal is to estimate the posterior probabilities \(\Pr(Y = j \mid \mathbf{X} = \mathbf{x})\) for each category (class) \(j\) for $j=1,\ldots, K$. Under a generative approach, suppose we have estimates \(\widehat\pi_j\) of the class priors and estimates \(\widehat f_{\mathbf{X}\mid Y=j}(\mathbf{x}\mid j)\) of the class-conditional joint density/pmf of \(\mathbf{X}\) given \(Y=j\). Then Bayes' theorem provides
\begin{equation}
\widehat{\Pr}\bigl(Y = j \mid \mathbf{X} = \mathbf{x}\bigr)
\;=\;
\frac{\widehat\pi_j \,\widehat f_{\mathbf{X}\mid Y=j}(\mathbf{x}\mid j)}
     {\displaystyle\sum_{k=1}^K \widehat\pi_k \,\widehat f_{\mathbf{X}\mid Y=k}(\mathbf{x}\mid k)},
\label{eq:bayes}
\end{equation}
which ensures \(\sum_{j=1}^K \widehat{\Pr}(Y = j \mid \mathbf{X} = \mathbf{x}) = 1\). Here \(\widehat f_{\mathbf{X}\mid Y=j}(\mathbf{x}\mid j)\) denotes the estimated joint density for continuous variables and/or probability mass for discrete variables of \(\mathbf{X}\) in class \(j\). In Equation \eqref{eq:bayes}, each class‐conditional density/pmf \(\widehat f_{\mathbf{X}\mid Y}(\mathbf{x}\mid j)\) can be modeled using vine-based distributions, leading to a \emph{vine copula-based classifier} as in \cite{sahin2024vine}. 
%\cite{sahin2024vine} empirically showed that vine-based classifiers outperform naive Bayes and normal discriminant analysis when the variables violate independence or Gaussian assumptions. They also observed that vine copula-based classifiers can be more robust to unbalanced classes than random forests, unless one applies under-/over-sampling. 
The implementation of vine copula-based classifiers is given in the \texttt{R} package \texttt{vineclass} \citep{vineclass}. 

\subsection{Probabilistic classification measures}
Since vine copula-based classifiers provide posterior probabilities, it is natural to evaluate their classification performance using probabilistic scores. Accordingly, \cite{sahin2024vine} proposed the negative log-likelihood (nll) score, defined by $
\mathrm{nll}
=
-\frac{1}{n}
\sum_{i : y_i = j}
\log\bigl(\hat p_{j}(\mathbf{x}_i)\bigr),
$
where $n$ is the total number of observations, \(\mathbf{x}_i\) is the \(i\)th variable vector, \(y_i\) its true class label in $\{0,1\}$, and \(\hat p_{j}(\mathbf{x}_i)\) is the predicted posterior probability for the class $j$. In this score, lower is better, reflecting better calibration and discrimination of the method. 

Here, we propose to use the negative log-likelihood score \emph{per class} to assess the classification performance in the presence of unbalanced data, meaning that one class has many more observations than the others. By focusing on each class separately, we can assess whether the classifier calibrates probabilities well for minority and majority classes:
\[
\mathrm{nll}_j
=
-\frac{1}{n_j}
\sum_{i : y_i = j}
\log\bigl(\hat p_{\,j}(\mathbf{x}_i)\bigr),
\]
where \(n_j=|\{i:y_i=j\}|\) with $|.|$ denoting the cardinality of a set.

Likewise, we adopt the Brier score  \citep{brier1950verification} \emph{per class}:
\[
\mathrm{Brier}_j
=
\frac{1}{n_j}
\sum_{i : y_i = j}
\Bigl[
(1 - \hat p_{\,j}(\mathbf{x}_i))^2
\Bigr].
\]
The Brier score per class also assesses performance on class \(j\).  Lower \(\mathrm{Brier}_j\) indicates more accurate and well‐calibrated probability estimates for that class.

\subsection{Illustration for comparison to logistic regression and random forests}
\begin{figure}[ht!]
    \centering
    \begin{subfigure}{\textwidth}
        \centering
               \includegraphics[width=1\linewidth]{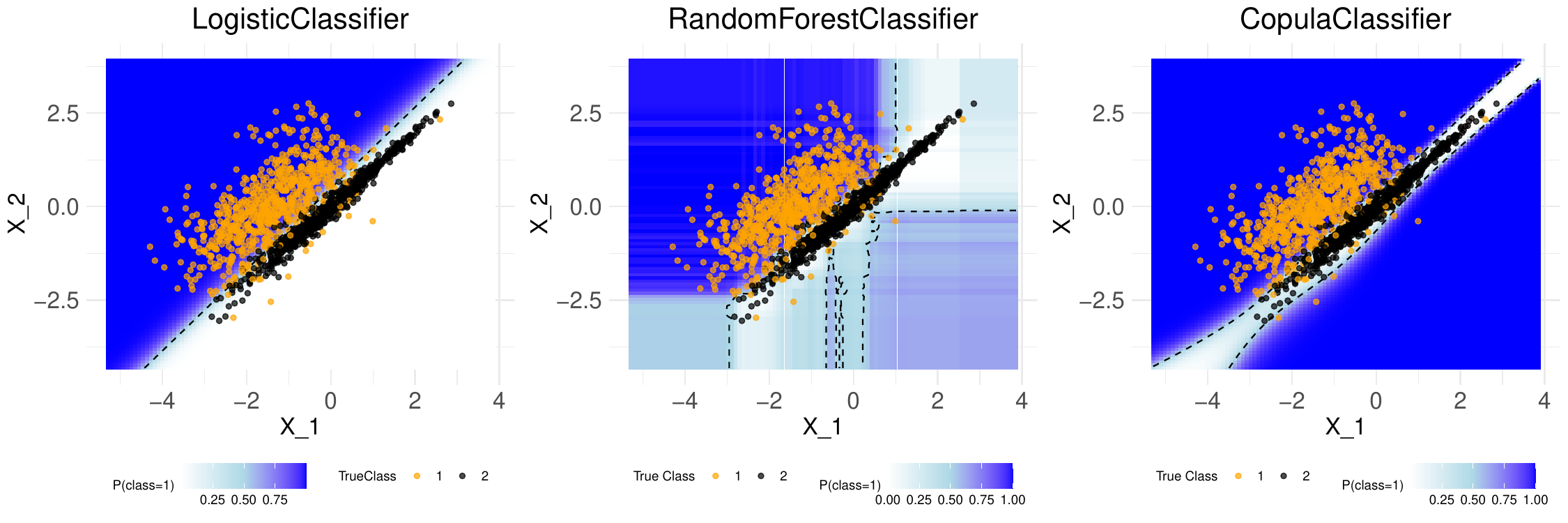}
        \caption{In-sample data.}
    \end{subfigure}
    \vspace{1cm}
    \begin{subfigure}{\textwidth}
        \centering
        \includegraphics[width=1\linewidth]{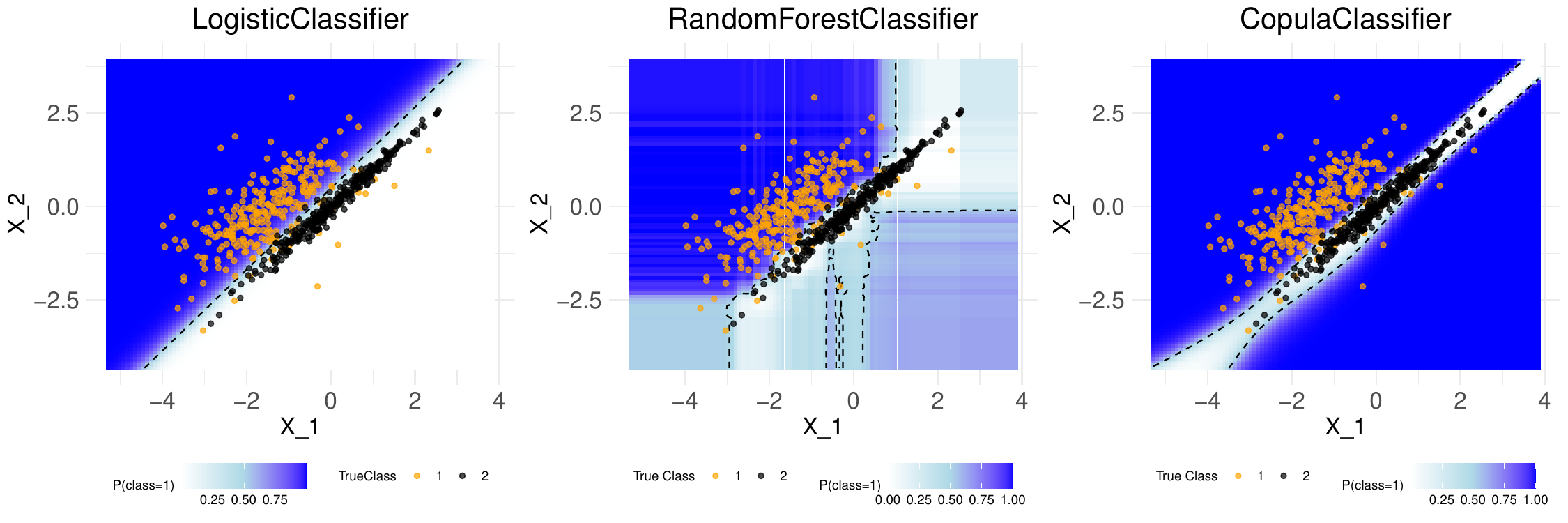}
        \caption{Out-of-sample data.}
    \end{subfigure}
    \caption{Binary classification decision boundaries shown by dashed lines at a 50\% probability threshold and predicted probabilities overlaid with contour colors for class 1 (orange)  for in-sample data (700 observations per class, above) using logistic regression, random forest, and copula models, based on a simulated data of two continuous variables specified in \autoref{sec:app-dgp}. The boundaries and the estimated probabilities are also shown with out-of-sample data (300 observations per class, below). The negative log-likelihood (nll) scores for each method are: logistic regression (in-sample: nll = 264.95, out-of-sample: nll = 142.61), random forest (in-sample: nll = 97.87, out-of-sample: nll = 126.08), and copula-based classifier (in-sample: nll = 177.28, out-of-sample: nll = 82.88). Lower nll values indicate a better model. Random forest uses 500 trees with at least 10 observations in leaf nodes. The copula-based classifier uses kernel-based density estimation for marginal densities and true copula families (in \autoref{sec:app-dgp}) in each class.}
    \label{fig:rflogcop-cont}
\end{figure}

%In Section \ref{sec:data-statcomp}, we will apply the vine copula-based classifier in a real clinical data set with unbalanced classes and compare its classification performance with logistic regression and random forests. 
As a visual illustration of copula-based classifiers, we consider a bivariate, binary classification scenario where data are generated from copulas (see \autoref{sec:app-dgp}). \autoref{fig:rflogcop-cont} shows the estimated classification decision boundaries by logistic regression, random forests, and a copula-based classifier, both in-sample and out-of-sample, along with overall nll scores. Even though the true data-generating process includes copulas, and thus the copula-based classifier is expected to be the best, \autoref{fig:rflogcop-cont} highlights key reasons empirically: (i) nonlinearity: logistic regression with default (linear) predictors does not capture interactions or nonlinear relationships; (ii) high variance of probability estimates in sparse regions: random forests can model nonlinear boundaries but often result in less confident probability predictions for sparse regions; (iii) copula flexibility: a correctly specified copula-based classifier can capture dependence patterns in tail and central regions, leading to more confident probabilistic assignments. 
%A similar pattern emerges in \autoref{fig:rflogcop-mixed}, where one variable is continuous and the other is ordinal. 

%\begin{figure}[ht!]
%    \centering
%    \begin{subfigure}{\textwidth}
%        \centering
%               \includegraphics[width=1\linewidth]{fig/logrfcop-train-mixed.pdf}
%        \caption{In-sample data.}
%    \end{subfigure}
%    \vspace{0.5cm}
%    \begin{subfigure}{\textwidth}
%        \centering
%        \includegraphics[width=1\linewidth]{fig/logrfcop-test-mixed.pdf}
%        \caption{Out-of-sample data.}
%    \end{subfigure}
%    \caption{Binary classification decision boundaries shown by dashed lines at a 50\% probability threshold and predicted probabilities overlaid with contour colors for class 1 (orange)  for in-sample data (700 observations per class, above) using logistic regression, random forest, and copula models, based on a simulated data of one continuous and one discrete variable specified in \autoref{sec:app-dgp}. The boundaries and the estimated probabilities are also shown with out-of-sample data (300 observations per class, below). The negative log-likelihood (nll) scores for each method are: logistic regression (in-sample: nll = 338.12, out-of-sample: nll = 169.13), random forest (in-sample: nll = 246.65, out-of-sample: nll = 148.43), and copula-based classifier (in-sample: nll = 258.29, out-of-sample: nll = 111.45). Lower nll values indicate a better model. Random forest uses 500 trees with at least 10 observations in leaf nodes. The copula-based classifier uses kernel-based density estimation for marginal densities and true copula families (in \autoref{sec:app-dgp}) in each class.}
%    \label{fig:rflogcop-mixed}
%\end{figure}

\subsection{Probabilistic patient risk profiling}\label{sec:riskgroups}
A common clinical question is, "How likely is it that a given patient will experience an adverse outcome?" For example, one might want to predict the probability (risk) of developing postoperative serious complications following bowel surgeries. Knowing such probabilities can guide medical decisions, such as shortened hospital stays and discharge with remote monitoring devices.

To aid these medical decisions, we propose a probabilistic approach that places each patient into a "risk group" based on the estimated probability of belonging to class $j^{*}$, where $j^\star\in\{1,\dots,K\}$ is the adverse outcome class like developing serious complications, given variables  $\bm{X}$, given by $\widehat{P}\bigl(J=j^{*}\mid \bm{X}=\bm{x}\bigr)$.  Here $\widehat P(. \mid .)$ is computed for each patient's variable vector $\bm x$, but we suppress the patient index for brevity. In particular, vine copula-based classifiers result in such posterior probabilities, but \emph{any} method returning posterior class probabilities would be applicable. 

Below, we illustrate one way to construct three risk groups (low, moderate, high) around two thresholds determined by a parameter $\alpha\in(0,1)$. Specifically:
\begin{itemize}
    \item \textbf{high-risk}: $\widehat{P}(J=j^{*}\mid\bm{X}=\bm{x}) \,\ge\, 1-\alpha$,
    \item \textbf{low-risk}: $\widehat{P}(J=j^{*}\mid\bm{X}=\bm{x}) \,\le\, \alpha$,
    \item \textbf{moderate-risk}: otherwise.
\end{itemize}
For instance, if $\alpha=0.25$ and $\widehat{P}(J=j^{*}\mid\bm{X}=\bm{x})=0.95$, the patient is placed into \emph{high-risk}. If that probability is $0.10$, the patient is \emph{low-risk}; and if it is $0.60$, the patient is \emph{moderate-risk}. High- or moderate-risk patients may be candidates for remote monitoring strategies after surgeries, while low-risk patients might be discharged early.

Here, a key challenge is choosing the threshold parameter $\alpha$. Generally, a higher $(1-\alpha)$ threshold leads to fewer high-risk patients but increases the confidence that these patients are truly at risk. In practice, threshold selection can be guided by historical data and cost-benefit analyses. For example, if hospital bed availability is limited, one might keep $\alpha$ smaller (thus reducing the high-risk group). %Likewise, if the cost or burden of continuous monitoring is high, one may want a smaller $\alpha$.

%Even if a patient is deemed high-risk, hospitalization alone may not prevent complications; additional clinical protocols (e.g., specialized nursing care or further testing) might be necessary for prevention. 

In Section \ref{sec:data}, we will illustrate how these ideas can be implemented using real clinical data. We will show how to derive risk groups based on $\widehat{P}(J=j^{*}\mid \bm{X}=\bm{x})$  vine copula-based classifier leads to and how the size and composition of each risk group changes with different thresholds.

Naturally, clinicians or hospital administrators must refine these risk definitions and thresholds according to logistical constraints, patient safety goals, and available interventions for high- and moderate-risk groups. In some scenarios, postoperative variables (e.g., time until mobility or early symptom scores) might also be integrated to update these risk profiles dynamically.

%In summary, by pairing vine copula-based classifiers (or any probabilistic classification method) with thresholds, one can obtain a straightforward patient risk profiling system to guide individualized discharge and monitoring decisions.

\section{Application: patient risk profiling with vine copula-based classifiers}\label{sec:data}

We analyze 767 elective bowel surgeries performed at Medisch Spectrum Twente (MST, Netherlands, 03/2020–12/2024) under the Enhanced Recovery After Surgery protocol \citep{Gustafsson2011}. The target is a binary outcome: a \emph{serious} complication if the Clavien–Dindo (CD) grade is $\ge$ IIIa, otherwise \emph{non-serious} \citep{clavien1992proposed}. Five predictors are retained from an initial set of 34 \citep{sahin2025}: three continuous (\texttt{BMI}, \texttt{surgerytime}, \texttt{age}) and two ordinal (\texttt{bloodloss}, \texttt{givencrystalloids}, each with three levels) to show ideas in low-dimensional space. Ethical approval was obtained from the MST non-WMO review board (K24-41, December 2024).

Data from March 2020 to December 2022 (518 non-serious, 62 serious) form the training set (in-sample); data from January 2023 to December 2023 (168/19) provide temporal validation (test; out-of-sample). Class imbalance motivates evaluation by class-specific Brier score and negative log-likelihood rather than AUC alone. We compare vine-copula classifiers with weighted logistic regression and stratified random forests (\texttt{randomForest}, 500 trees, min-leaf 10) \citep{RF}. All models run in under a minute in \texttt{R} 4.3.1 on an Apple M2 laptop.

Vine copula models are selected via the Gaussian-vine strategy of \cite{brechmann2015truncation}. Pearson, polyserial, and polychoric estimates give the latent correlation matrix; successive tree levels maximize summed (partial) correlations. Pair-copula families are chosen by the modified Bayesian Information Criterion of \cite{nagler2019model} from a candidate pool consisting of independence, Gaussian, Student-$t$, Clayton, Gumbel, Frank, and Joe copulas. We use equal class priors to prevent bias toward the majority class.

The remainder of Section \ref{sec:data} covers: exploratory analysis (Section \ref{sec:data-eda}); performance comparison (Section \ref{sec:data-statcomp}); interpretation and diagnostics of fitted vines (Section \ref{sec:data-interpretvine}–Section \ref{sec:data-diagnosvine}); risk-group identification (Section \ref{sec:data-identifyrisk}); and scenario analyses for clinical policy (Section \ref{sec:data-scenario}).

\subsection{Exploratory data analysis}\label{sec:data-eda}
Table~\ref{tab:eda_summary} compares the five predictors between surgeries with serious complications ($62$ observations) and those without ($518$ observations). We observe that the average mean of \texttt{age} and \texttt{BMI} is close across both classes. However, a higher proportion of patients in the serious complication class experienced high blood loss (42.0\% vs.\ 14.1\% in category 3) and received large volumes of crystalloids (32.3\% vs.\ 9.6\% in category 3). The average surgery duration was longer in the serious complication class than in the other class. \autoref{fig:hist} in Appendix \ref{sec:app_eda} shows the histograms of continuous variables, where we observe that patients with lower BMI may be more likely to experience complications.

\begin{table}[ht!]
\centering
\caption{Summary statistics of the variables by the complication class (serious vs. non-serious). For continuous variables, means and ranges are reported; for ordinal variables, percentages per category are shown.  Blood loss has three categories as "No loss (1)," "$\leq 100$ mL (2)," and "$>100$ mL (3)". Crystalloids volume administered has three categories as  "$< 1000$ mL, (1)" "$ \geq1000 \text{ and } < 2000$ mL (2)," and "$\geq2000$ mL (3)".}
\label{tab:eda_summary}
\scalebox{0.75}{
\begin{tabular}{llll}
\hline
Variable & Description & Non-serious & Serious \\
\hline
\texttt{age} & Age of patient & 65.4 [20--92] & 64.3 [24--88] \\
\texttt{BMI} & Body mass index & 26.4 [14.03--48.33] & 25.9 [14.47--39.79] \\
\texttt{bloodloss} & Blood loss & 1: 63.1\%, 2: 22.8\%, 3: 14.1\% & 
1: 41.9\%, 2: 16.1\%, 3: 42.0\% \\
 & (1: no, 2: moderate, 3: high) & \\
\texttt{givencrystalloids} & Crystalloids administered  &  1: 39.4\%, 2: 51.0\%, 3: 9.6\% &
1: 22.6\%, 2: 45.2\%, 3: 32.3\%\\
 &(1: low, 2: moderate, 3: high) &   \\
\texttt{surgerytime} & Surgery duration (min) & 114.6 [15--475] & 148.1 [32--383] \\
\hline
\end{tabular}}
\end{table}

\begin{figure}[ht!]
    \centering
        \caption{A heatmap of the (latent) correlation matrix for the in-sample data, where upper and lower triangular parts show estimated correlations among variables associated with serious (62 observations) and non-serious complications (518 observations), respectively. Diagonal values of 1 are omitted for visualization purposes. Continuous-continuous, continuous-discrete, and discrete-discrete correlations are estimated using the normal scores' Pearson, polyserial, and polychoric correlations, respectively.}
    \label{fig:corr-train}
    \includegraphics[width=.4\linewidth]{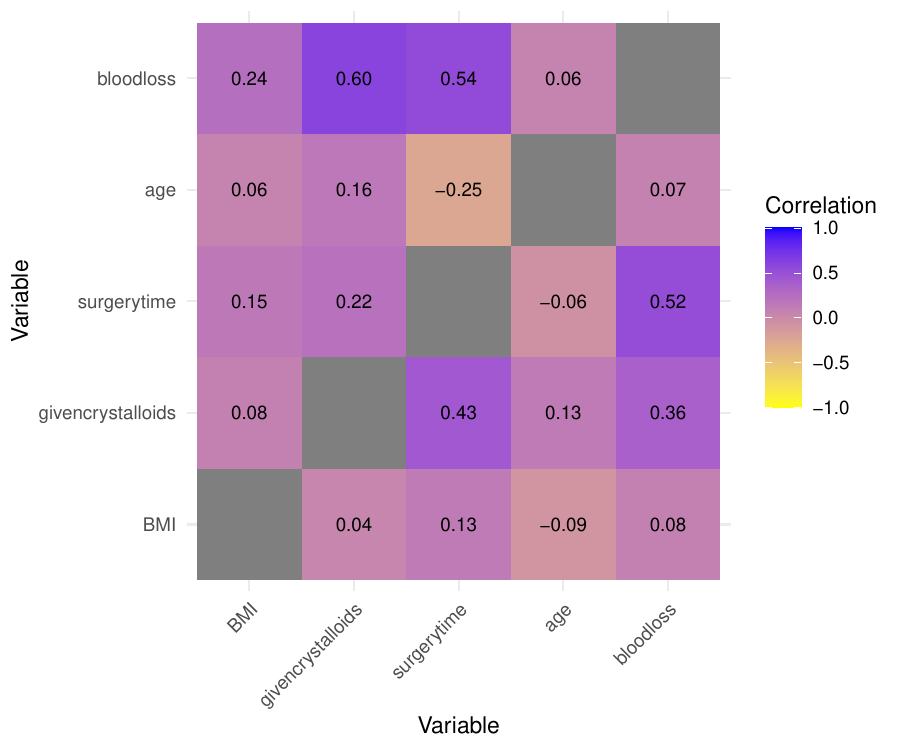} 
\end{figure}

\autoref{fig:corr-train} shows a heatmap of the estimated (latent) correlation matrix, separately for patients with serious (upper triangular) and non-serious (lower triangular) complications. Depending on the variable type, these estimates are based on polychoric, polyserial, and Pearson correlations. 

We observe notable patterns in \autoref{fig:corr-train}: first, \texttt{bloodloss} shows a moderate positive correlation with \texttt{surgerytime} in both complication classes (0.52 and 0.54). This might reflect an intuitive clinical relationship that longer surgeries are typically associated with higher blood loss. Second, the estimated association between \texttt{bloodloss} and \texttt{givencrystalloids} is weak in the non-serious complication class but becomes moderate in the serious complication class (0.36 vs. 0.60).  It might indicate the increased crystalloid administration needs in surgical cases with high bleeding.  The estimated correlation between \texttt{surgerytime} and \texttt{givencrystalloids} is higher in the non-serious complication class (0.43) compared to the serious complication class (0.22). It suggests that crystalloid administration during surgery tends to be more aligned with surgical duration in routine cases without serious complications than in those that result in serious complications. Lastly, the estimated correlation between \texttt{age} and \texttt{surgerytime} is negative across complication classes (-0.25 vs. -0.06), with the absolute value larger in the serious complication class. Older age may be linked to shorter surgical durations due to more conservative surgery decisions for older people.

\subsection{Classification performance comparison}\label{sec:data-statcomp}
We next evaluate the discrimination and calibration performance of vine copula-based classifiers compared to random forest with stratified sampling and weighted logistic regression (estimated coefficients are given in \autoref{sec:app-risk}) on the explored data. As discussed in Section~\ref{sec:vineclass}, due to data being unbalanced, we assess their performance using Brier and negative log-likelihood (nll) scores per class.

 The vine copula-based classifier achieves the most balanced performance across classes and in- and out-of-sample data out of three methods in \autoref{tab:brier_nll_scores}. In the non-serious complication class, the vine copula-based classifier has a Brier score of 0.19 in the in-sample data and 0.20 in the out-of-sample data. Likewise, its Brier score for the serious complication class is 0.22 and 0.23 for the in- and out-of-sample data, respectively. In addition, it has the lowest out-of-sample Brier (0.23) and nll (0.64) scores for the serious complication class.

Even though the random forest achieves the best in-sample Brier and nll scores for serious complications, its performance decreases in the out-of-sample data, potentially due to overfitting. While further hyperparameter tuning may improve its performance, such refinements fall outside the scope of this study. Likewise, the weighted logistic regression model underperforms in the serious complication class, particularly in the out-of-sample data. We will discuss potential reasons for these classification performance differences across models in the next subsection.

Even though the area under the receiver operating characteristic curve (AUC) is a widely used summary measure for discrimination, it is insensitive to calibration and does not measure class-specific performance of the model. 
%In our case study, the vine copula-based classifier, weighted logistic regression, and random forest with stratified sampling achieved AUCs of 0.67, 0.68, and 0.71, respectively, in the in-sample; corresponding values on the out-of-sample were 0.71, 0.60, and 0.66. 
One would rank the random forest highest based solely on the in-sample AUC, despite its poor out-of-sample performance. This highlights the importance of complementing discrimination metrics with probability-based scores such as the class-specific Brier and negative log-likelihood, as in \autoref{tab:brier_nll_scores}. 

\begin{table}[ht!]
\centering
\caption{Brier, negative log-likelihood (nll) scores and AUC per class (serious vs. non-serious complications) for the in-sample (518 non-serious and 62 serious) and out-of-sample (168 non-serious and 19 serious) data for the vine copula-based classifier, random forest with stratified sampling and weighted logistic regression. The best method per class and score is highlighted for the in-sample and out-of-sample data.}
\scalebox{0.85}{
\begin{tabular}{llccccc}
\hline
 && AUC & \multicolumn{2}{c}{Brier score} & \multicolumn{2}{c}{nll score} \\
\hline
 && & Non-serious & Serious & Non-serious & Serious \\
 \hline
\multirow{3}{*}{In-sample} 
 & Vine copula classifiers       & 0.67  & 0.19 & 0.22 & 0.57 & 0.62 \\
 & Random forest (strata) & \hl{0.81}& \hl{0.17} & \hl{0.10} & \hl{0.53} & \hl{0.35} \\
 & Weighted logistic reg. & 0.68& 0.21 & 0.22 & 0.62 & 0.63 \\ \hline
\multirow{3}{*}{Out-of-sample} 
 & Vine copula   classifiers       &  \hl{0.71} & \hl{0.20} & \hl{0.23} & 0.61 & \hl{0.64} \\
 & Random forest (strata) & 0.68& \hl{0.20} & 0.25 & 0.59 & 0.71 \\
 & Weighted logistic reg. &0.60 & \hl{0.20} & 0.29 & \hl{0.58} & 0.78 \\ \hline

\end{tabular}}
\label{tab:brier_nll_scores}
\end{table}

\subsection{Interpretation of the fitted vine copulas}\label{sec:data-interpretvine}  
Next, we will discuss how to interpret the fitted vine copula-based classifiers. %We fitted separate vine copulas for the serious and non-serious complication classes and selected pair-copula families and the truncation level via modified BIC as explained in Section~\ref{sec:vineclass}.  As a result, 
The fitted vine copula for the serious complication class is 1-truncated, meaning that only unconditional dependencies between pairs of variables are modeled. In contrast, the fitted vine copula for the non-serious complication class is 2-truncated. Hence, it models unconditional dependence and conditional dependence conditioning on only one variable. The difference in the truncation level of the vine copulas among the classes is likely driven by the small number of observations in the serious complication class (62 observations), for which early truncation also reduces the risk of overfitting.

\begin{figure}[ht!]
\caption{The first tree level of the estimated vine copula model for serious and non-serious complication classes in the in-sample data (518 with non-serious complications, 62 with serious complications). A letter at an edge with numbers inside the parenthesis refers to its bivariate copula family with its estimated parameter, where C: Clayton, F: Frank, G: Gumbel, J: Joe, N: Gaussian, SC: Survival Clayton, and RC270: $270^\circ$ rotated Clayton copula. Variable indices are \texttt{1:BMI}, \texttt{2:givencrystalloids}, \texttt{3:surgerytime}, \texttt{4:age}, and \texttt{5:bloodloss}. }
	\centering
	\renewcommand{\xshiftNodes}{0.1*\linewidth}
	\renewcommand{\yshiftLabels}{.0cm}  
\begin{tikzpicture}	[every node/.style = VineNode, node distance =1.6cm,scale=0.75, transform shape]
\node (v5)   {2}		
node             (v2)         [right of = v5, xshift = \xshiftNodes] {5}
node             (v1)         [right of = v2, xshift = \xshiftNodes] {1}
node             (v3)         [below of = v1] {4}			
node             (v4)         [left of = v3, xshift =- \xshiftNodes] {3}
node[TreeLabels] (T1)        [below of = v5] {Serious} ;
\draw[color=black] (v5) to node[draw=none,  font = \labelsize,fill = none, above, yshift = \yshiftLabels] {J(2.5)} (v2);
\draw[color=black] (v2) to node[draw=none,  font = \labelsize,fill = none, above, yshift = \yshiftLabels] {C(0.36)} (v1);
\draw[color=black] (v2) to node[draw=none, font = \labelsize, fill = none, left, xshift = 0.5cm, yshift = \yshiftLabels] {SC(1.3)} (v4);
\draw[color=black] (v3) to node[draw=none, font = \labelsize, fill = none, below, yshift = \yshiftLabels] {RC270(0.40)} (v4);
\node (v5f)   [right of = v1, xshift=3*\xshiftTree]     {5}			
node             (v2f)         [right of = v5f, xshift = \xshiftNodes] {3}
node             (v1f)         [right of = v2f, xshift = \xshiftNodes] {1}
node             (v3f)         [below of = v2f] {2}			
node             (v4f)         [right of = v3f, xshift = \xshiftNodes] {4}
node[TreeLabels] (T2)        [below of =v5f] {Non-serious};
\draw[color=black] (v5f) to node[draw=none,  font = \labelsize,fill = none, above, yshift = \yshiftLabels] {G(1.4)} (v2f);
\draw[color=black] (v2f) to node[draw=none,  font = \labelsize,fill = none, above, yshift = \yshiftLabels] {F(0.84)} (v1f);
\draw[color=black] (v2f) to node[draw=none,font = \labelsize, fill = none,left, xshift = 0.5cm, yshift = \yshiftLabels] {N(0.43)} (v3f);
\draw[color=black] (v3f) to node[draw=none, font = \labelsize, fill = none, above, yshift = \yshiftLabels] {N(0.12)} (v4f);	
\end{tikzpicture}
\label{fig:vine_structure}
\end{figure}
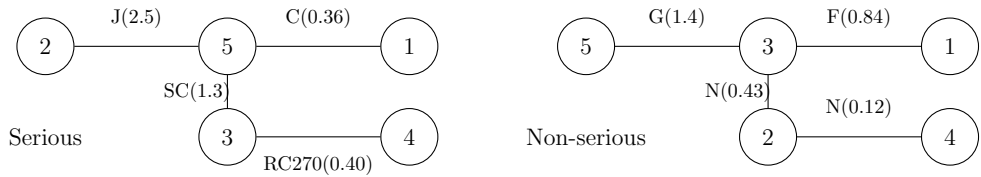

~\autoref{fig:vine_structure} shows the first tree level of the estimated vine copula models. We observe that in the non-serious complication class, most pairwise dependence (3 out of 4) are modeled by Gaussian or Frank copulas, thereby implying symmetric dependence among the modeled pairs of variables (e.g., \texttt{surgerytime} and \texttt{givencrystalloids}). On the other hand, none of the pair copulas in the vine tree of the serious complication class is symmetric but tail-asymmetric. For example, the pairwise dependence between \texttt{givencrystalloids} and \texttt{bloodloss} is modeled by a Joe copula, reflecting the upper tail dependence. This implies that patients experiencing extreme blood loss also tend to receive more crystalloid administration.

In the second tree of the fitted vine copula of the non-serious complication class (more details are available upon request), \texttt{BMI} and \texttt{givencrystalloids} are modeled as conditionally independent once \texttt{surgerytime} is taken into account. This suggests that after controlling for \texttt{surgerytime}, \texttt{BMI} does not explain additional variation in \texttt{givencrystalloids}. Further, the estimated conditional Kendall's tau (Spearman's rho) between \texttt{surgerytime} and \texttt{age} conditioned on \texttt{givencrystalloids} is -0.07 (-0.11) for the non-serious complication class. Due to being a 1-truncated vine, the second tree of the fitted vine copula of the serious complication class consists of independence copulas. It implies that conditional dependencies are negligible once the asymmetric pairwise dependence in the first tree level is accounted for.

%As shown in \autoref{tab:brier_nll_scores}, the vine copula-based classifier outperforms weighted logistic regression in discrimination and calibration metrics, particularly for the serious complication class. Unlike logistic regression, which assumes additive linear effects, vine copulas flexibly capture nonlinear and tail-asymmetric dependence as shown in \autoref{fig:vine_structure}. Further, they can model conditional independence through the independence copula whenever needed and offer parsimony and interpretability.  

\subsection{Diagnostics for conditional dependence and pair-copula fit}\label{sec:data-diagnosvine}
Next, we focus on diagnostic tools for assessing the conditional dependence as a function of the conditioning variable (evaluating the simplifying assumption) and checking pair-copula fit adequacy from bivariate margins for the vine copulas fitted to the serious and non-serious complication classes. These diagnostics help evaluate whether more flexible structures, such as non-simplified vines or alternative pair-copula families, might be needed for a better model fit.

While tests of the simplifying assumption have been primarily developed \citep{kurz2022testing, derumigny2023testing, nagler2025simplified}, \cite{chang2020copula} propose to use conditional Spearman's rho and corresponding confidence bands to analyze if the conditional dependence is constant over the conditioning value. We simply extend this approach to the case where the conditioning variable is ordinal. Because conditioning is on an ordinal variable, no smoothing or bandwidth selection is required. We assess sampling variability using a non-parametric bootstrap, with 90\% confidence bands computed from 1000 replicates.

\begin{figure}[ht!]
    \centering
    \caption{Conditional Spearman's rho between \texttt{surgerytime} and \texttt{age} across the three categories of \texttt{givencrystalloids} in the non-serious complication class. Black points indicate the observed conditional Spearman's rho per category, with 90\% bootstrap confidence intervals. The red dashed line shows the modeled conditional dependence from the fitted vine copula.}
    \label{fig:conditional_rho}
    \includegraphics[width=0.4\linewidth]{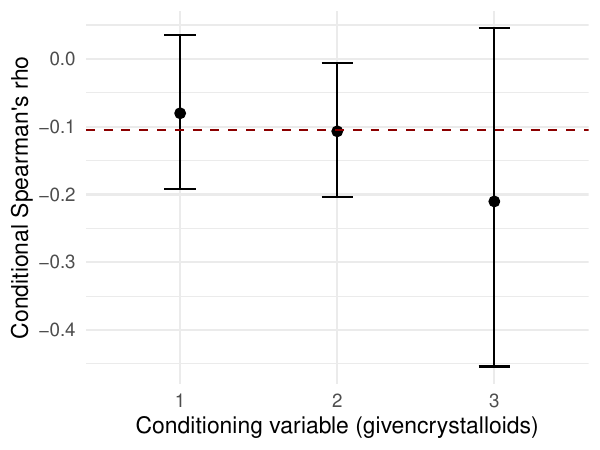}
\end{figure}

As detailed in the previous section, the vine copula fitted to the serious complication class is 1-truncated and therefore does not include conditional dependence. Thus, no diagnostics of conditional dependence are needed for that class. In contrast, the vine copula for the non-serious complication class models the conditional dependence between \texttt{surgerytime} and \texttt{age} given \texttt{givencrystalloids}. To assess this conditional relationship, we estimate the conditional Spearman's rho between \texttt{surgerytime} and \texttt{age} for each category of the ordinal variable \texttt{givencrystalloids} as detailed above.  As shown in \autoref{fig:conditional_rho}, the modeled conditional Spearman’s rho lies within the bootstrap confidence bands across all categories. Hence, we find no empirical indication of a simplifying assumption violation for this edge.

Next, we assess bivariate copula fits using normal score plots, which visually highlight tail asymmetry and deviations from a Gaussian copula. We adopt the latent normal score construction introduced by \cite{pan2024assessing} for continuous-ordinal pairs. This method replaces the ordinal variable with a pseudo-continuous latent Gaussian variable, preserving its empirical marginal distribution and the polyserial correlation with the continuous variable.

\begin{figure}[ht!]
    \centering
        \caption{Normal score plot of \texttt{surgerytime} and the latent Gaussian variable generated from \texttt{bloodloss} in the non-serious complications class (left) and that of \texttt{surgerytime} and \texttt{age} in the serious complications class (right).}
    \label{fig:normal_score}
    \includegraphics[width=0.75\linewidth]{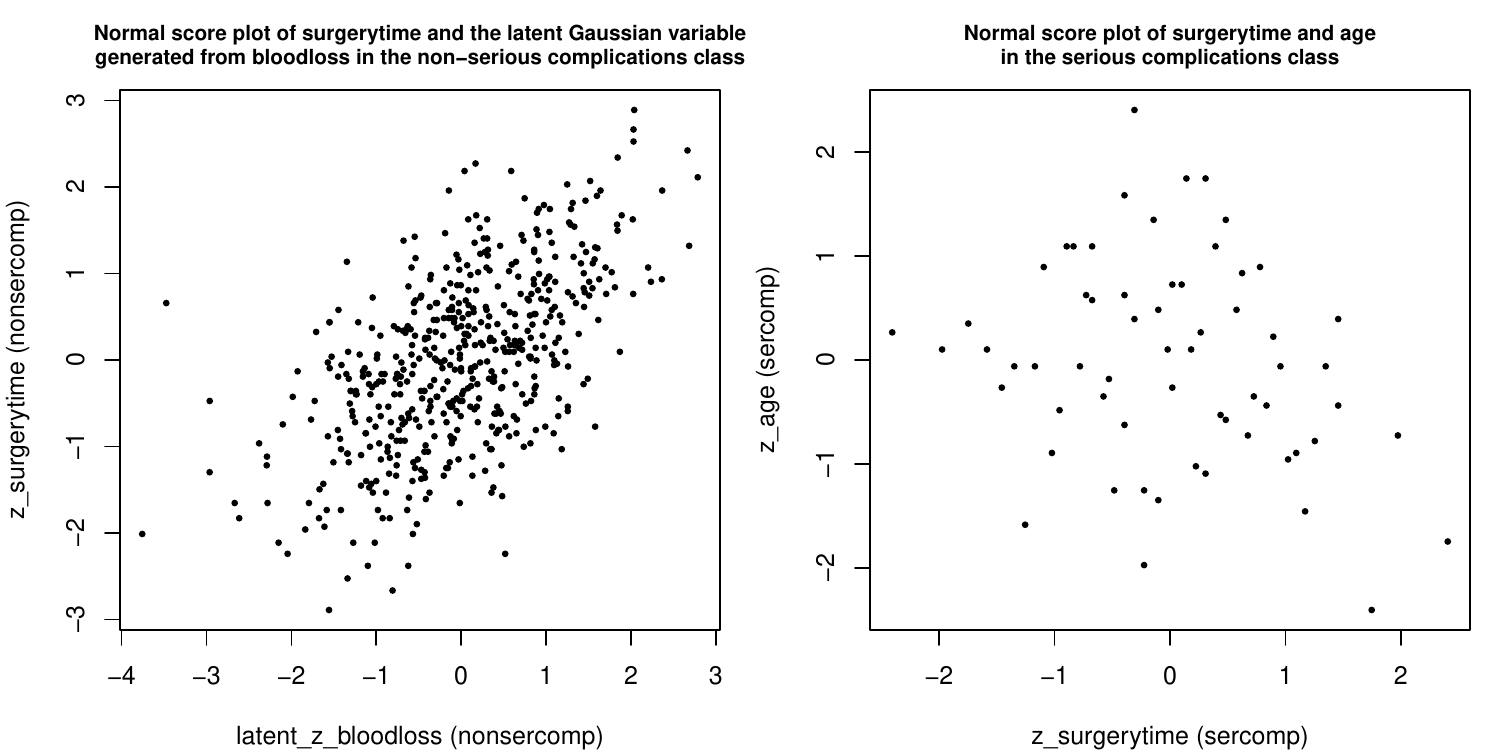}
\end{figure}

The left panel of \autoref{fig:normal_score}  shows the normal score plot of \texttt{surgerytime} and the latent Gaussian version of \texttt{bloodloss} in the non-serious complication class. Stronger upper-tail dependence than expected under bivariate normality is visible. Therefore, it supports the fitted Gumbel copula that captures upper tail dependence (see \autoref{fig:vine_structure}). In the right panel, the plot for \texttt{surgerytime} and \texttt{age} in the serious complication class exhibits stronger right lower-tail dependence, which is also consistent with the rotated Clayton copula used in the vine copula model (see \autoref{fig:vine_structure}).

Additional normal score plots for other variable pairs (continuous-continuous and continuous-ordinal) are available upon request. Overall, our diagnostics support the adequacy of the fitted pair-copulas. We refer interested readers to \cite{nikoloulopoulos2015factor} for diagnostic methods in the discrete-discrete pair of variables.

\subsection{Identification of patient risk groups}\label{sec:data-identifyrisk}
Given that diagnostics support the adequacy of pair-copula fits, in this section, we identify three distinct patient risk groups (low-risk, moderate-risk, and high-risk) based on the estimated posterior probabilities of (not) developing serious complications given by the vine copula-based classifier across varying thresholds $\alpha$ (see Section~\ref{sec:riskgroups} for the methodology). 

\autoref{tab:riskgroups} summarizes these risk groups for the in-sample and out-of-sample data. The good discrimination and calibration of the vine copula-based classifiers reported in \autoref{tab:brier_nll_scores} are reflected in the patient risk group's identification in  \autoref{tab:riskgroups}. For instance, at $\alpha=0.15$, a patient is assigned to the low-risk group if their posterior probability of not developing a serious complication is at least 0.85. Under this threshold, the vine copula-based classifier effectively identifies patients who indeed did not develop serious complications: 26 in-sample and 10 out-of-sample patients are in the low-risk group with no observed serious complications in both cases. This group corresponds to the shortest average hospital stays, with 3.08 nights in the in-sample and 4.40 nights in the out-of-sample data.

In \autoref{tab:riskgroups}, most patients are placed into the moderate-risk group; this also reflects that the discrimination across both classes is hard. While the high-risk group is accurately identified in the in-sample data with a higher average length of stay and a significant proportion of serious complications, its identification is less successful in the out-of-sample data, as seen in the smaller number of patients with their serious complication status.

As seen in \autoref{tab:riskgroups}, increasing the threshold $\alpha$ expands the low-risk and high-risk groups, as expected. For example, at $\alpha=0.25$, the low-risk group expands to 92 patients in the in-sample data and 29 patients in the out-of-sample data. However, this increase comes at the cost of misclassification: one out-of-sample patient and two in-sample patients in the low-risk group developed a serious complication. These results emphasize the importance of carefully selecting $\alpha$ based on the desired balance between identifying truly low-risk patients and minimizing false negatives. 

The identification of a reliable low-risk group offers clear opportunities for health economic optimization. Patients in the low-risk group, who did not experience serious complications after surgeries, could benefit from early discharge protocols. For example, the average length of stay can be safely reduced to two nights from the current average stay of three nights for the low-risk group. As a result, it might free up resources and reduce healthcare costs without compromising patient safety.

As discussed in Section~\ref{sec:riskgroups}, our proposal for patient risk groups can be applied using any classification method that estimates posterior class probabilities. In Appendix~\ref{sec:app-risk}, we report the risk groups obtained from the weighted logistic regression at the thresholds $\alpha=0.15$ and $\alpha=0.25$. At $\alpha=0.25$, its low-risk group includes a patient with serious complications and fewer patients without serious complications than the vine copula-based classifier (16 vs. 92 in the in-sample data and 6 vs. 29 in the out-of-sample data). Further, at $\alpha=0.15$, it cannot identify any patients for the low-risk group, while the vine-based approach places 26 and 10 patients into the low-risk group in the in-sample and out-of-sample, respectively. In addition, the vine copula-based classifier had a stricter separation between low- and high-risk patients. This makes the vine-based approach more appealing for health economic optimization in this case study.

As discussed in Section~\ref{sec:riskgroups}, our proposal for patient risk groups can be applied using any classifier that estimates posterior probabilities. In Table \ref{tab:riskgroups}, we compare results from weighted logistic regression and stratified random forests at $\alpha = 0.15$ and $0.25$. Both methods identify broader moderate-risk groups and are less conservative in assigning low-risk patients. For instance, at $\alpha=0.25$, logistic regression places only 16 in-sample and 6 out-of-sample patients in the low-risk group (vs.\ 92 and 29 for the vine classifier), while random forests identify 168 and 44, respectively, with slightly higher misclassification. In contrast, the vine classifier maintains a cleaner separation between risk strata, especially in terms of length of stay. 

\begin{table}[ht!]
\centering
\caption{True serious complication classes (True class: 0 = non-serious complications, 1 = serious complications) and average hospital nights after surgery (Avg night: average with estimated standard deviation in parentheses) across risk groups estimated by the vine copula-based classifier, weighted logistic regression, and random forests with strafied sampling under different $\alpha$ thresholds defined in Section \ref{sec:riskgroups} for the in-sample (580 observations) and out-of-sample data (187 observations).}
\label{tab:riskgroups}
\scalebox{0.75}{
\begin{tabular}{|l|l|rr|r|rr|r|}
\hline
\multicolumn{8}{|c|}{Vine copula-based classifier} \\ \hline
$\alpha$ & Risk group & \multicolumn{3}{|c|}{In-sample} & \multicolumn{3}{|c|}{Out-of-sample} \\ \cline{3-8}
& & \multicolumn{2}{|c|}{True class} & Avg night & \multicolumn{2}{|c|}{True class} & Avg night \\ \hline
& & 0 & 1 & &  0 & 1 & \\ \hline
0.15 & Low-risk        & 26  & 0  & 3.08 (1.94) & 10  & 0  & 4.40 (2.63) \\
     & Moderate-risk  & 480 & 51 & 5.61 (7.10) & 155 & 18 & 7.24 (7.45) \\
     & High-risk      & 12  & 11 & 12.50 (9.62) & 3   & 1  & 12.50 (8.81) \\ \hline
0.20 & Low-risk        & 56  & 1  & 3.65 (4.38) & 18  & 0  & 4.22 (2.37) \\
     & Moderate-risk  & 438 & 41 & 5.35 (6.75) & 141 & 18 & 7.42 (7.70) \\
     & High-risk      & 24  & 20 & 13.10 (10.30) & 9 & 1  & 9.10 (6.33) \\ \hline
0.25 & Low-risk        & 92  & 2  & 3.71 (4.08) & 29  & 1  & 4.70 (3.25) \\
     & Moderate-risk  & 392 & 38 & 5.43 (6.99) & 124 & 16 & 7.68 (8.06) \\
     & High-risk      & 34  & 22 & 11.90 (9.60) & 15  & 2  & 7.65 (5.45) \\ \hline
\multicolumn{8}{|c|}{Weighted logistic regression} \\ \hline
$\alpha$ & Risk group & \multicolumn{3}{|c|}{In-sample} & \multicolumn{3}{|c|}{Out-of-sample} \\ \cline{3-8}
& & \multicolumn{2}{|c|}{True class} & Avg night & \multicolumn{2}{|c|}{True class} & Avg night \\ \hline
& & 0 & 1 & &  0 & 1 & \\ \hline
0.15 
     & Moderate-risk  & 508 & 55 & 5.56 (7.07) & 167 & 18 & 7.15 (7.32) \\
     & High-risk      & 10  & 7 & 12.60 (8.74) & 1   & 1  & 11.50 (12.0) \\ \hline
0.25 &Low-risk        & 16  & 0  & 3.19 (1.97)  & 6   & 1  & 8.14 (9.28) \\
 &Moderate-risk  & 470 & 47 & 5.38 (6.96)  & 158 & 15 & 6.91 (7.15) \\
 &High-risk      & 32  & 15 & 10.90 (8.85) & 4   & 3  & 13.40 (8.38) \\ \hline

\multicolumn{8}{|c|}{Stratified random forest} \\ \hline
$\alpha$ & Risk group & \multicolumn{2}{|c|}{True class} & Avg night & \multicolumn{2}{|c|}{True class} & Avg night \\ \cline{3-8}
& & 0 & 1 & &  0 & 1 & \\ \hline
0.15 & Low-risk        & 51  & 0  & 3.24 (2.09) & 7   & 1  & 5.88 (5.33) \\
     & Moderate-risk   & 463 & 47 & 5.73 (7.29) & 160 & 17 & 7.21 (7.40) \\
     & High-risk       & 4   & 15 & 13.60 (8.74) & 1   & 1  & 11.50 (12.0) \\ \hline
0.25 & Low-risk        & 168 & 0  & 3.70 (2.78) & 44  & 1  & 5.84 (7.39) \\
     & Moderate-risk   & 315 & 30 & 5.37 (6.37) & 110 & 15 & 7.34 (7.46) \\
     & High-risk       & 35  & 32 & 13.00 (12.7) & 14  & 3  & 9.71 (5.76) \\ \hline

\end{tabular}
}
\end{table}

\subsection{Scenario-based risk profiling}\label{sec:data-scenario}
The fitted vine copulas for the serious and non-serious complication classes enable scenario-based analyses: by varying two variables while holding others fixed at clinically relevant values, we construct two-dimensional risk surfaces that visualize how the predicted serious complication probability evolves across the variable space.

\begin{figure}[ht!]
    \centering
    \caption{Estimated posterior probabilities of serious complications across the joint space of \texttt{BMI} and \texttt{surgerytime}, for fixed variables \texttt{givencrystalloids} = 2 and \texttt{age} = 65. Left panel: no blood loss (\texttt{bloodloss} = 1); right panel: high blood loss (\texttt{bloodloss} = 3). \texttt{BMI} and \texttt{surgerytime} were varied on equally spaced grids from 15 to 35 and 30 to 180, respectively. A black contour line shows the 0.50 probability threshold for binary classification (if present).}
    \label{fig:heatmap}
    \includegraphics[width=.45\linewidth]{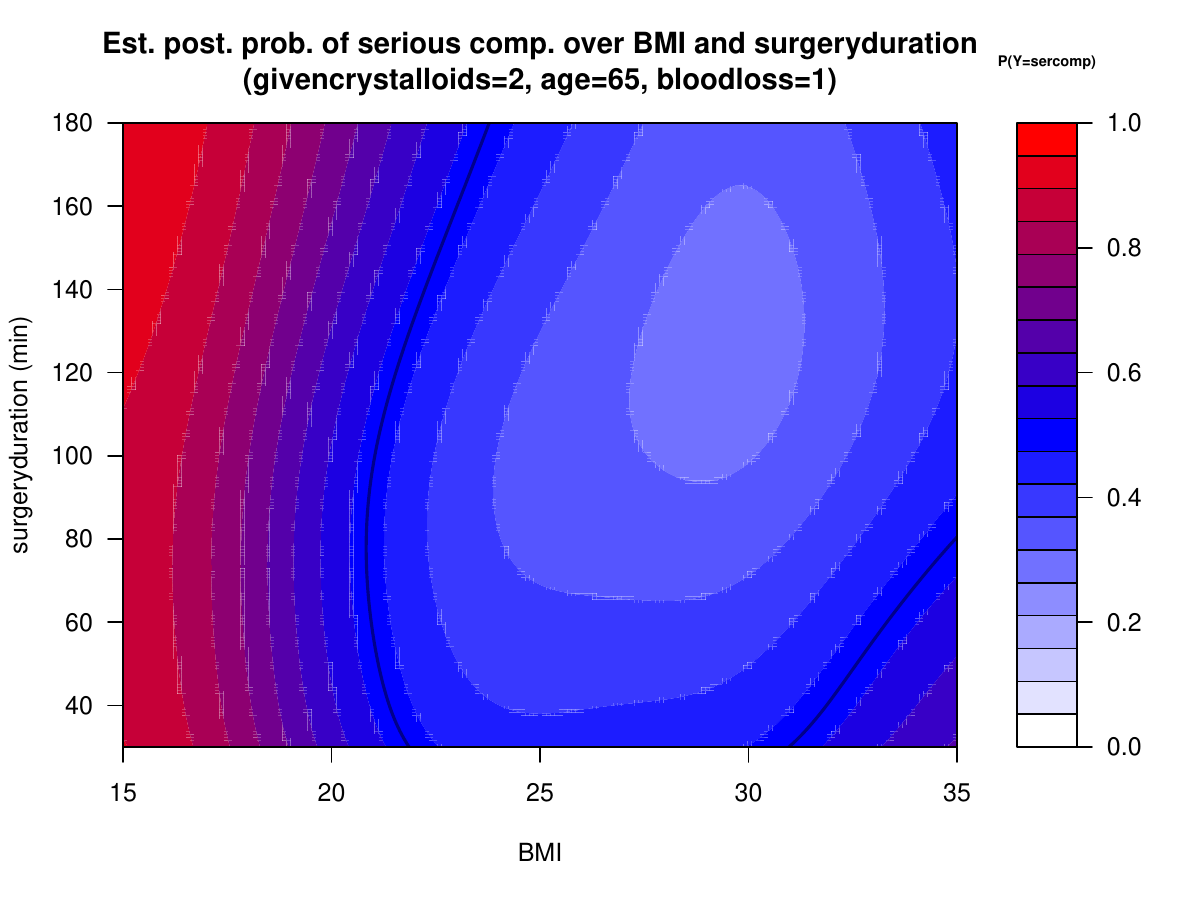} 
    \includegraphics[width=.45\linewidth]{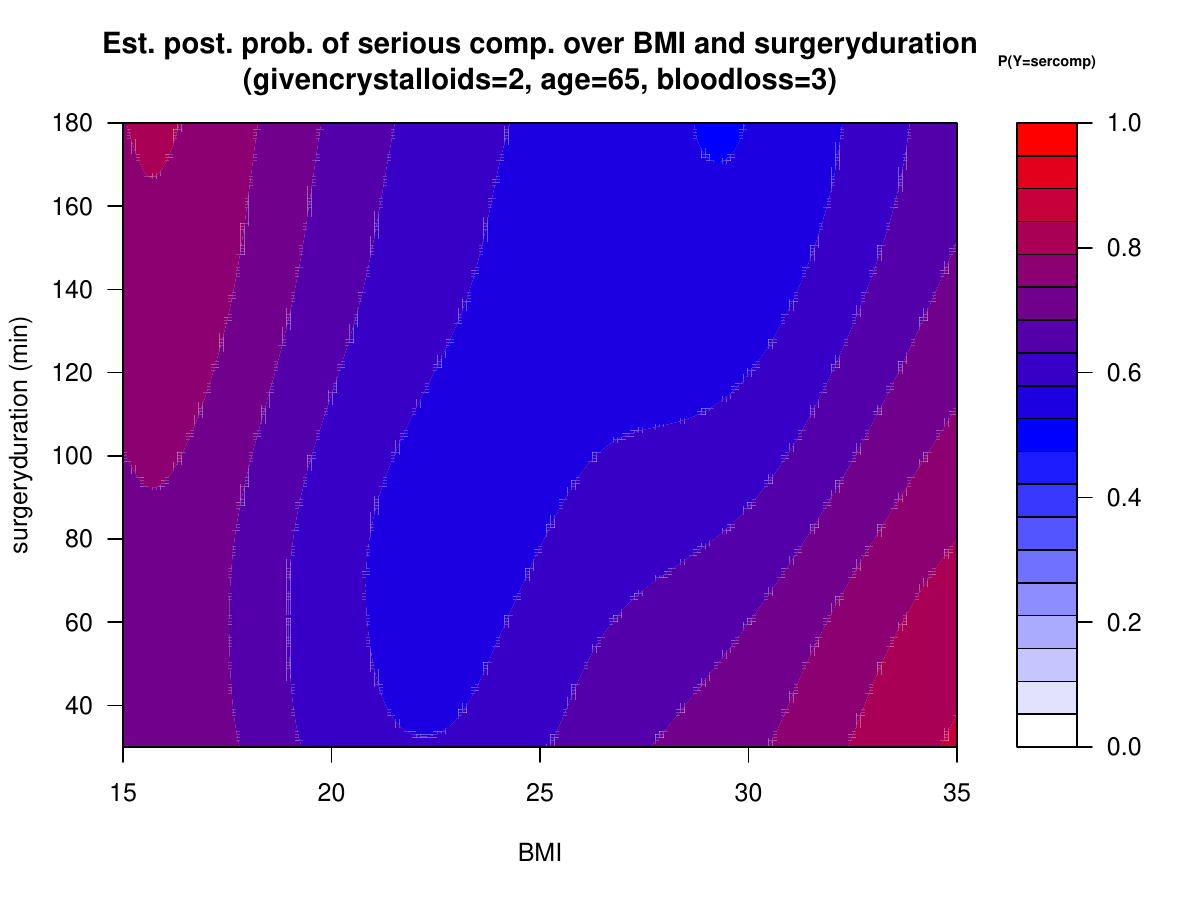} 
\end{figure}

As an example, we analyze how the surgery duration and \texttt{BMI} impact the predicted probabilities for developing serious complications jointly. \autoref{fig:heatmap} presents estimated risk surfaces over \texttt{BMI} and \texttt{surgerytime}, under two blood loss scenarios: no (\texttt{bloodloss} = 1, left panel) and high (\texttt{bloodloss} = 3, right panel). The other two variables were fixed at \texttt{age} = 65 and \texttt{givencrystalloids} = 2 to reflect a typical surgery in our case study. First, the estimated risk surface highlights a nonlinear interaction between \texttt{BMI} and surgery duration, where risk does not follow simple additive patterns. 

In the left panel of \autoref{fig:heatmap}, which represents no blood loss, we observe a broad safe zone for the serious complication risk, indicated by light (blue) colors, where the estimated probabilities for serious complications are lower than 0.25. It spans approximately \texttt{BMI} between 24 and 32 and \texttt{surgerytime} between 75 and 180. Further, we see a black contour line marking the 0.50 probability threshold. Still, patients with a \texttt{BMI} below 18.5 (underweight) have a high risk of developing serious complications (the corresponding estimated probability is higher than 0.75) regardless of their surgery duration. 

In contrast, the right panel of \autoref{fig:heatmap} illustrates the high blood loss scenario. Here, the predicted probability of serious complications exceeds 0.50 throughout the plotted domain; thus, no contour line appears. This shows that, for a typical 65-year-old patient with moderate crystalloid volume, experiencing high blood loss leads to the probability estimate of developing serious complications higher than 0.50, irrespective of \texttt{BMI} or surgery time. Still, patients with approximately normal weight (21-24.9) and shorter surgeries ($\leq$90) or those approximately with the \texttt{BMI} range of 25-28 and a surgery duration range of 110-160 have the estimated probabilities for serious complications under 0.60. 
%In the previous subsection, we observed that for a normal-weight, 65-year-old patient (\texttt{BMI} = 25) undergoing a two-hour surgery with moderate crystalloid volume, blood loss did not push the estimated probability of serious complications beyond 0.57. However, the joint analysis in \autoref{fig:heatmap} provides a more comprehensive analysis: blood loss is a dominant factor of serious complication risks when interacting with a broader range of \texttt{BMI} and \texttt{surgerytime}  values.

\section{Discussion and conclusion}\label{sec:disc}
Vine copulas capture nonlinear and tail asymmetric dependence structures between mixed-type (continuous-ordinal) clinical variables. Combined with Bayes' theorem, the resulting generative classifier, vine copula-based classifiers, provides nonlinear classification boundaries observed in clinical data and posterior probabilities, enabling patient-specific risk profiling.

In this paper, we demonstrated the benefits of vine copula-based classifiers for probabilistic patient risk profiling. Our case study focused on postoperative outcomes following elective bowel surgeries, with a binary outcome defined as serious versus non-serious complications based on the Clavien-Dindo classification. %Serious complications were defined as CD grade~$\ge$IIIa, while non-serious ones included either no complication or CD grade~$<$IIIa. 
The training set consisted of 62 surgeries in the serious complication class and 518 in the non-serious class; the test set included 19 and 168 observations, respectively. In our analyses, vine copula-based classifiers achieved better class-specific Brier and negative log-likelihood scores compared to weighted logistic regression and random forests with stratified sampling, in both in-sample and out-of-sample. Empirically, we show that these differences can result from three properties of vine copulas: the ability to flexibly model nonlinear relationships, explicit handling of tail dependencies, and model selection approaches that allow for sparsity. We further confirmed the adequacy of the fitted pair-copulas from the vine copula-based classifiers with the proposed diagnostic tools for pair-copulas.

We proposed a patient risk profiling approach based on posterior probabilities from any classifier, including vine copula-based classifiers. By selecting a threshold for the posterior probabilities, we group patients into low-, moderate-, and high-risk categories. Our case study showed that the vine copula-based classifiers identified more patients in the low-risk group than weighted logistic regression, with no false negatives at the same threshold level. Reliable identification of such a low-risk patient group with vine copula-based classifiers can support health economic optimization, for example, by reducing the number of hospital stays after elective bowel surgeries.

We also presented scenario-based risk analyses using fitted vine copulas from vine copula-based classifiers. We showed how changes in body mass index and surgery time impact the probability of developing serious complications for a 65-year-old patient with no blood loss and moderate crystalloid administration via nonlinear risk surfaces. Clinically, such risk surfaces can guide targeted preoperative strategies, such as offering nutritional support to underweight patients or allocating closer monitoring for patients with high blood loss after elective bowel surgeries.

Despite the promising findings, our study has limitations. The number of patients in the serious complication class is relatively small, which may limit the copula model's ability to detect strong tail dependencies in some variable pairs. Moreover, even though data collection protocols remained unchanged, part of our data includes the COVID-19 period, which may have introduced temporal variation. Given the limited number of observations, our analyses do not explicitly account for such heterogeneity.

Future research directions include developing variable selection strategies for vine copula-based classifiers with mixed-type variables. Sensitivity analyses concerning vine structure, pair-copula families, and comparison with Gaussian or independence assumptions can also improve model understanding. Since the outcome is ordinal, one can alternatively model it as a discrete response in a vine copula regression framework to estimate posterior probabilities directly, similar to multinomial or ordinal logistic regression. Future work will systematically compare vine copula regression and vine copula-based classifiers, with guidelines for selecting the appropriate method depending on sample size, outcome type, and clinical setting. 

From an applied point of view, future work includes incorporating postoperative variables such as nausea presence or the days until mobility to refine our risk groups with clinicians' discharge decisions. Validation on larger and external datasets would also be needed to evaluate transportability.

\subsection*{Acknowledgement}
We thank Reini Bretveld for providing access to the surgical dataset. Grammarly was used for language editing; all scientific content is the authors' own.

\subsection*{Statements and declarations}

\textbf{Funding}: This work is part of the 4TU programme RECENTRE (Risk-based lifEstyle Change: daily-lifE moNiToring and REcommendations), funded by the 4TU programme High Tech for a Sustainable Future. 4TU is the federation of the four technical universities in the Netherlands (Delft University of Technology, Eindhoven University of Technology, University of Twente, Wageningen University and Research). 

\textbf{Declaration of conflicting interest}: All authors have none to declare.

\textbf{Data availability}: Due to confidentiality, data are not publicly available but can be requested from the corresponding author, subject to Medisch Spectrum Twente approval.

\newpage
\section*{Appendix}
\appendix

%\section{Samples generated from copulas coupled with margins}\label{sec:app-copulas}

%\begin{figure}[H]
%    \centering
 %       \caption{1000 observations generated from Gumbel copula with the %parameter corresponding to Kendall's $\tau$ of 0.75 coupled with different margins specified in captions. Each point in the plots has a size proportional to the number of observations with that value.}
 %   \includegraphics[width=1.0\linewidth]{fig/gumbelCop1.pdf}

 %   \label{fig:gumbelCop}
%\end{figure}

%\begin{figure}[H]
%    \centering
%        \caption{1000 observations generated from Clayton copula with the parameter corresponding to Kendall's $\tau$ of 0.75 coupled with different margins specified in captions. Each point in the plots has a size proportional to the number of observations with that value.}
%    \includegraphics[width=1.0\linewidth]{fig/claytonCop1.pdf}
%
%    \label{fig:claytonCop}
%\end{figure}

%\begin{figure}[H]
%    \centering
%    \includegraphics[width=1.0\linewidth]{fig/gaussianCop1.pdf}
%    \caption{1000 observations generated from Gaussian copula with the parameter corresponding to Kendall's $\tau$ of 0.75 coupled with different margins specified in captions. Each point in the plots has a size proportional to the number of observations with that value.}
%    \label{fig:gausssianCop}
%\end{figure}

\section{Data generating process for simulated data}\label{sec:app-dgp}
The data-generating process (DPG) for the simulated data shown in \autoref{fig:rflogcop-cont} is as follows:

\begin{enumerate}
    \item Specify copula models: Class 1 ($j=1$): Frank copula with Kendall's tau of 0.5 (\( \tau = 0.5 \)). 
    Class 2 ($j=2$): Gumbel copula with Kendall's tau of 0.9 (\( \tau = 0.9 \)).

    \item Sample 1000 observations for each class $j$ on the copula scale, \( (\bm{U} | j=1) \) and \( (\bm{U}|j=2) \),  using the respective copula models, where \( (\bm{U} | j=1) \) =\( (U_1 | j=1 , U_2 | j=1 )^\top\).
    
    \item Transform the variables to the x-scale:
    \[
   ( X_1 | j=1) = \mathcal{N}^{-1}((U_1 | j=1), \mu_1, \sigma) \quad (X_2 | j=1) = \mathcal{N}^{-1}((U_2 | j=1), \mu_y, \sigma)
    \]
    \[
    (X_1 | j=2) = \mathcal{N}^{-1}((U_1 | j=2), \mu_2, \sigma) \quad (X_2 | j=2) = \mathcal{N}^{-1}((U_2 | j=2), \mu_y, \sigma)
    \]
    with \(\mu_1 = -1.5\), \(\mu_2 = 0\), \(\mu_y = 0\), and \(\sigma = 1\), and $\mathcal{N}^{-1}(.)$ corresponds to the inverse cumulative distribution function of the univariate normal distribution.

\end{enumerate}

%The DGP for the simulated data shown in \autoref{fig:rflogcop-mixed} is as follows:

%\begin{enumerate}
%    \item The same as the DGP of \autoref{fig:rflogcop-cont} given above.

%    \item The same as the DGP of \autoref{fig:rflogcop-cont} given above.
    
%    \item Transform the variables to the x-scale:
%    \[
%   ( X_1 | j=1) = \mathcal{N}^{-1}((U_1 | j=1), \mu_1, \sigma) \quad (X_2 | j=1) = \text{Poisson}^{-1}((U_2 | j=1), \lambda_y)
%    \]
 %   \[
%    (X_1 | j=2) = \mathcal{N}^{-1}((U_1 | j=2), \mu_2, \sigma) \quad (X_2 | j=2) = \text{Poisson}^{-1}((U_2 | j=2), \lambda_y)
%    \]
%    with \(\mu_1 = -1.5\), \(\mu_2 = 0\),  \(\sigma = 1\), and \(\lambda_y = 2\).  $\mathcal{N}^{-1}(.)$ and $\text{Poisson}^{-1}(.)$ correspond to the inverse cumulative distribution function of the univariate normal distribution and the Poisson distribution, respectively. 

%\end{enumerate}

\section{Data}\label{sec:app_eda}

\begin{figure}[H]
    \centering
        \caption{Histograms of \texttt{BMI}, \texttt{surgerytime}, and \texttt{age} grouped by complication status, with patients with serious complications (sercomp) to those without (nonsercomp).}
         \label{fig:hist}
    \includegraphics[width=\linewidth]{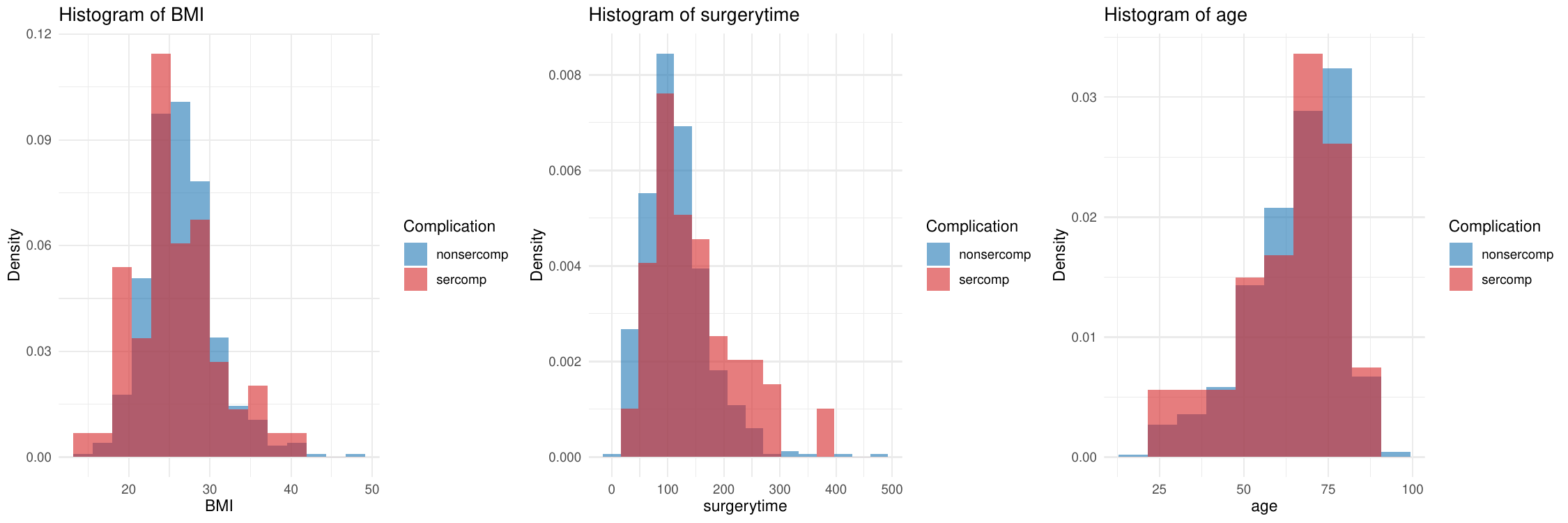}
\end{figure}

\section{Weighted logistic regression results}\label{sec:app-risk}
Estimated coefficients from the weighted logistic regression, where weights for each observation are inversely proportional to their respective class frequency, are as follows: \texttt{(Intercept)} = 0.591, \texttt{BMI} = -0.050, \texttt{cryst.=2} = 0.321,  \texttt{cryst.=3} =1.127,  \texttt{surgtime} =0.004, \texttt{age} = -0.006, \texttt{bloodl.=2} = -0.055,  \texttt{bloodl.=3} = 0.863.
\raggedright
%\printbibliography

\bibliography{vineclassHealth}
\end{document}